\documentclass[12pt]{article}
\usepackage{graphicx}
\begin{document}
\baselineskip = 12pt
\newcommand{\etal}{{\it etal.\/}}

\title{Influence of dopant concentration on the structure and physical properties of Nd$_{1-x}$Pb$_{x}$MnO$_{3}$ single crystals }
\author{
Nilotpal Ghosh$^{a}$, Suja Elizabeth$^{a}$, H.L. Bhat$^{a}$\footnote{corresponding 
author. fax: 91-80-360 2602, E-mail: hlbhat@physics.iisc.ernet.in},\\
 G. Nalini$^{b}$, B.Muktha$^{b}$ and T.N. Guru Row$^{b}$\\{\small $^{a}$Department of Physics, Indian Institute of Science, Bangalore-560012,
India}\\{\small $^{b}$Solid State and Structural Chemistry Unit, Indian Institute of Science, Bangalore-560012, India
 }}
\date{}
\maketitle


\begin{abstract}
\baselineskip = 18pt

The  structure of Nd$_{1-x}$Pb$_{x}$MnO$_{3}$ crystals is determined by single crystal X-ray diffraction. Substitution of Pb at the Nd site results in structural phase change from  tetragonal (x = 0.25) to cubic (x = 0.37).
These changes  are attributed  to the progressive removal of inter-octahedral tilting and  minimization of the octahedral distortion leading to a higher symmetry as doping concentration increases. While the unit cell volume of tetragonal structure ($\it{P4/mmm}$) is comparable to that of parent NdMnO$_{3}$, the volume of cubic unit cell ($\it{ Pm\overline{3}m}$) is doubled. Electron diffraction patterns support these results and rule out the possibility of twinning. Changes in transport properties as a function of temperature at different doping levels are in accordance with the observed structural changes. It is noticed that $\it{T_{c}}$ and $\it{T_{MI}}$ increase with x.

\end{abstract}

\section*{ Introduction}


Rare-earth manganites exhibiting colossal magnetoresistance (CMR) continue to attract research attention  because of diverse physical properties and possible technological applications. Subtle structural changes  between multiple phases and their relationship with physical properties have led to detailed structural investigations. The RMnO$_{3}$ perovskites (R is any rare earth between Lanthanam and Dysprosium) are orthorhombically distorted ($\it{Pnma}$) and isostructural with GdFeO$_{3}^{\ref{gill},\ref{woo}}$. On the other hand, the crystal structure is hexagonal with space group $\it{P6_{3}cm}$ for rare-earths from Holmium to Lutetium as well as Yttrium$^{\ref{woo},\ref{poll}}$. Substitution of trivalent lanthanides  by divalent alkaline earth ions (Ca, Sr, Ba and Pb) leads to the  simultaneous occurrence of Mn$^{3+}$  and Mn$^{4+}$  in the crystal lattice. This modifies the structural and physical properties significantly.
A number of structural modifications have been discovered in manganite systems$^{\ref{pg}-\ref{jt}}$. In general,  majority of CMR compounds fall into two types: rhombohedral ($\it{R\overline{3}m}$) and orthorhombic ($\it{Pnma/Pbnm}$). 
A structural change is noticed in manganites with  increase in divalent doping. 
For  example, the structure of La$_{1-x}$Sr$_{x}$MnO$_{3}$  changes from orthorhombic ($\it{Pnma}$) to rhombohedral ($\it{R\overline{3}m}$) as doping  concentration increases$^{\ref{jf},\ref{Urushi}}$. However, other symmetries are also observed as in the case of  Nd$_{1-x}$Sr$_{x}$MnO$_{3}$ which changes  from  orthorhombic ($\it{Imma}$) to tetragonal ($\it{I4/mcm}$)  system as x increases$^{\ref{kaji}}$. 
 Symmetry relationships between perovskite polymorphs have been investigated from a group theoretical standpoint. This demonstrates a path of continuous phase transition through a sequence of space group-subgroup steps which is consistent with experiments$^{\ref{noel}}$.\\

The parent compounds  (LaMnO$_{3}$, NdMnO$_{3}$ and PrMnO$_{3}$) are A-type antiferromagnetic insulators. Their insulating nature as well as anisotropic magnetic interaction are related to structure. There are two characteristic distortions that influence the perovskite  structure of manganites. One relates to co-operative tilting  of MnO$_{6}$ octahedra which is  established below 1000K  as  a consequence of  mismatch of ionic radii between the rare-earth and divalent ions$^{\ref{rao}}$. This often takes the form of rigid rotations of the MnO$_{6}$ octahedra and modifies the co-ordination sphere of the R-site cation$^{\ref{tok}}$. The other, Jahn-Teller (JT) distortion arises due to electronic instability of Mn$^{3+}$ in the octahedral crystal field  influenceing the MnO$_{6}$ octahedra in such a way that there are long and short Mn-O bonds and deviations in O-Mn-O bond angles from 180$^{0}$.
These two distortions play a key role in structural evolution of perovskite manganites.
Electronic transport is often strongly influenced by  structural changes. For example, in La$_{1-x}$Sr$_{x}$MnO$_{3}$ the  rhombohedral phase is metallic while the orthorhombic phase is insulating  at low temperature $^{\ref{jf},\ref{Urushi}}$ where as in La$_{1-x}$Ca$_{x}$MnO$_{3}$, the orthorhombic symmetry is retained both in insulating and metallic phases. Local structural features  have profound influence on the conduction mechanisms in manganites $^{\ref{tok}}$. As divalent ions are doped in these systems, the ferromagnetic (FM) double exchange (DE)  Mn$^{3+}$-O- Mn$^{4+}$ interaction begins to compete with antiferromagnetic (AFM) super exchange (SE) Mn$^{3+}$-O-Mn$^{3+}$  interaction. The basic mechanisms of SE and DE interactions  are  mediated  by  the  d-electron overlap between two transition metal ions via the intervening oxygen p-orbital. The  extent of  overlap depends on cation-oxygen-cation bond angle and cation-cation bond length in different perovskite systems$^{\ref{jb}-\ref{kana}}$. In manganites, this overlap integral  depends on Mn-O-Mn bond angle and Mn-Mn bond  length  which are sensitive to the tilt of neighbouring  MnO$_{6}$ octahedra $^{\ref {jf},\ref{coey}}$. 
 Hence it is interesting to study the influence of divalent dopant on the local structural modifications and the relevant physical properties. This paper reports the study of Pb substitution in NdMnO$_{3}$  and its impact on the structure, transport and magnetic properties.

\section*{ Crystal Growth}
 Nd$_{1-x}$Pb$_{x}$MnO$_{3}$  single crystals are obtained by high temperature solution growth technique using PbO/PbF$_{2}$  solvent with Nd$_{2}$O$_{3}$, MnCO$_{3}$ and PbO  in stoichiometric quantities as solute $^{\ref{nil}}$.  
The homogenized mixture was contained in a platinum crucible and growth was carried out in a resistive furnace. The cationic ratio was varied to obtain a range of x values. The composition of resulting crystals was determined by Energy Dispersive X-ray (EDX) Analysis followed by Inductively Coupled Plasma Emission Spectroscopy (ICPAES) for better accuracy.  

 \section*{Structural Studies}
A crystal of nominal lead content x = 0.25 and size 0.3x0.2x0.04 mm$^{3}$  was mounted on a BRUKER AXS SMART APEX CCD diffractometer $^{\ref{bruk}}$ at a distance of 6.03 cm from the detector. The diffraction intensities were measured with monochromated MoK$_{\alpha}$ radiation ($\it{\lambda}$ = 0.7107 $\AA$). The orientation matrix was obtained at room temperature from reflections derived from 50 frames to give the tetragonal unit cell (Table 1). Data were collected in four batches covering a complete sphere of reciprocal space with each batch at different $\it{\phi}$ angles ($\it{ \phi}$ = 0, 90, 180, 270$^{0}$ ) and each frame covering 0.3 degree in $\it{\omega}$ at 10 seconds exposure time. The data was 98.7$\%$  complete to 56$^{0}$ in 2$\it{\theta}$. An approximate empirical absorption correction was applied assuming the crystal shape to be cylindrical. Positional coordinates of Nd and Mn atoms were obtained by direct methods using the SHELX97 module and refined by SHELXL97 $^{\ref{shel}}$. 
 The Nd atoms occupy three crystallographically distinct sites (2e, 1d and 1b). In order to account for the presence of Pb in any of Nd sites, occupancy refinements were carried out (constraining the overall site occupancy to 1.0) indicating that the Pb atom fully occupies the 1d site and amounts to the presence of 25$\%$ of Pb at Nd sites. Hence, refinement confirms that Pb and Nd occupy different crystallographic sites (1d and 2e,1b respectively)  suggesting ordering at R-site in RMnO$_{3}$ perovskite structure$^{\ref{faal}}$.
 Difference Fourier synthesis revealed the positions of the remaining oxygen atoms, which  were included in the refinement.
 The occupancy refinements on the oxygen atoms suggest an occupancy of 96$\%$ for O(1) and 80$\%$ for O(3) leading to an acceptable R index of 8.3$\%$ (Table 2).
The structure of crystals with higher lead content (x = 0.4) was determined in a similar way. A crystal of size 0.18x0.057x0.2mm$^{3}$ was used for data collection. The orientation matrix was obtained from reflections derived from 400 frames at 12 seconds exposure time giving a cubic unit cell.
The data was 97.9$\%$ complete to 55$^{0}$ in 2$\it{\theta}$.
An empirical absorption correction was applied assuming the crystal shape to be nearly spherical. The positional coordinates of Pb/Nd and Mn atoms were obtained as before. Occupancy refinements carried out to ascertain the amount of Pb and Nd resulted in a ratio of 57:37 (Nd:Pb) with 96$\%$ Nd(2) at 3c site and 72$\%$ Nd(3) at 1b site. According to results of refinements it is observed again that Nd and Pb occupy different crystallographic  sites (1a, 3c and 1b for Nd and 3d for Pb)$^{\ref{faal}}$.
 The  occupancy refinements  of oxygen atoms indicated full occupancy in their respective sites (Table 2). 
The results  of measurement and structural refinement are summarized in Table 1. The atomic coordinates along with the equivalent thermal parameters are listed in Tables 2 and 3. The packing diagrams of tetragonal and cubic structures are shown in Fig.1(a) and (b). The structural refinement data obtained from single crystal x-ray diffraction experiments are used to simulate the powder x-ray diffraction pattern for both the crystals. Fig.2(a) and (b) show that they are matching well with the experimental powder diffratograms.\\
In order to rule out the possibility of twinning, electron diffraction experiments are carried out. Fig.3(a) and (b) show no evidence of twinning
for both the tetragonal and cubic phases. Superlattice spots are seen  along $\it{a}$ direction for tetragonal and along $\it{a}$ and $\it{c}$ directions for cubic crystal. Hence, the significance of cell doubling is quite evident for the cubic phase as shown in Fig.3(b).

\section*{ Physical Properties}

Transport and magnetization measurements were carried out on crystals of various doping levels. Resistivity was measured by four probe method in the temperature range 300 to 80 K and shown in Fig.4. Magnetic properties were determined by A.C susceptibility measurements in the same temperature range. All the doped crystals exhibited a paramagnetic to ferromagnetic transition as the temperature is lowered. Transition temperature T$_{c}$  progressively shifted upwards with increase in x values as seen in Fig. 5. 

\section*{Results and Discussions}
 The parent NdMnO$_{3}$ has an inherent perovskite distortion  due to the presence of Nd in  the R-site. As Nd is partially replaced by the larger cation Pb, the distortion is progressively reduced and the symmetry becomes tetragonal at 25$\%$ substitution. The distortion is almost removed in the case of 37$\%$ lead substitution which results in cubic symmetry as well as cell doubling. It is also evident from the electron diffraction pattern (Fig.3(b)).
The unit cell parameters can be related as a$_{t}$ = b$_{t}$ $\approx$ 2a$_{p}$; c$_{t}$ $\approx$ a$_{p}$ for tetragonal structure and a$_{c}$ $\approx$ 2a$_{p}$ for cubic structure where a$_{p}$ is cell parameter
of ideal primitive ($\it{Pm\overline{3}m}$) perovskite structure. The unit cell consists of corner
sharing regular MnO$_{6}$ octahedra in which Mn-O bonds become more symmetric as the structure changes from tetragonal to cubic (Tables 4 and 5). However the Pb-O bonds are significantly shorter than the Nd-O bonds in both structures. All Mn-O-Mn bond angles are nearly equal (178$^{0}$(3)) in cubic structure while they are dissimilar in the tetragonal phase. Consequently the  co-operative tilting of MnO$_{6}$ octahedra observed in NdMnO$_{3}$ is reduced considerably  in  Nd$_{0.75}$Pb$_{0.25}$MnO$_{2.72}$ and the least in Nd$_{0.57}$Pb$_{0.37}$MnO$_{3}$. R-site cationic deficiency is noticed in cubic crystals as seen in the formula (after refinement) while tetragonal crystals are oxygen deficient. The oxygen deficiency is known to occur in manganites$^{\ref{hl}}$ while cationic deficiency observed in the cubic structure is rare. The apparent R-site deficiency may be attributed to the refinement result which is strongly affected  by excessive X-ray absorption due to higher lead content in the cubic crystal.\\
 
Aleksandrov$^{\ref{alek}}$ and Barnighausen$^{\ref{bar}}$ investigated
the  existence of symmetry relationship between perovskite polymorphs from the view point of group theory. Aleksandrov reports transition routes from $\it{Pm\overline{3}m}$ to $\it{Pnma}$ symmetry via  five intermediate space groups$^{\ref{noel}}$. On the other hand, Barnighausen demonstrated path of continuous phase
transitions from $\it{Pm\overline{3}m}$ to $\it{Pnma}$ symmetry 
via a sequence of space group-subgroup steps $\it{Pm\overline{3}m-P4/mmm-Cmmm-Icmm-Pbnm(Pnma)}^{\ref{noel}}$. The structural transformation route observed in the present investigation is in accordance with  Barnighausen's  prediction, although intermediate stages like $\it{Cmmm}$ and $\it{Icmm}$ were not observed. These intermediate structures might be discernible
at other doping concentrations.\\

Fig.6 shows the variation of different Mn-O bond lengths with divalent dopant concentration. The static distortion of MnO$_{6}$ octahedra is maximum for parent orthorhombic  NdMnO$_{3}$(x = 0)$^{\ref{Mun}}$. The mismatch between different Mn-O bond lengths  of Nd$_{1-x}$Pb$_{x}$MnO$_{3}$ is much less at x = 0.25 and 0.37 for the tetragonal  and cubic structures respectively. In manganites, $\left<r_{A}\right>$ gives the measure of one electron (e$_{g}$) band width$^{\ref{tok}}$ which is very sensitive to the concentration and ionic radii of divalent dopant. It is noted that when $\left<r_{A}\right>$ $\gg$ 1.31 $\AA$ (x = 0.2) the Mn-O bond lengths are almost equal (inset of Fig. 6). The distortion in perovskite unit cell  consequent to  its  octahedral tilting  is  estimated by the tolerance  factor ($\it{t_{G}}$).
The structural change of Nd$_{1-x}$Pb$_{x}$MnO$_{3}$  from orthorhombic to cubic
(via tetragonal) with increase in x is  consistent with  the $\it{t_{G}}$ approaching unity. The dependence of $\it{t_{G}}$ on  dopant concentration (x) is shown in Fig.7.\\

 It is seen from the temperature dependence of resistivity (Fig.4) that the metal-insulator transition temperature ($\it{T_{MI}}$) (vertical arrows in the figure) increases with x.  
The magnitude of resistivity at room temperature decreases with increase of  x. This is  attributed to the  increment of  Mn-O-Mn bond angle with x (or $\left<r_{A}\right>$) which results in increase of Mn-Mn electron hopping rate$^{\ref{hy}}$.  
The plot for sample with x = 0.2 has a plateau-like region around 127 K suggesting  the onset of metallic phase. However the sample exhibits  an insulating behaviour as temperature is further lowered. But, the  resistivity profile for sample with x = 0.3  flattens  after 146 K as temperature decreases indicating metal like behaviour. Thus, one can see that the MI transition starts to occur when lead concentration is 20\% (x = 0.2) in Nd$_{1-x}$Pb$_{x}$MnO$_{3}$  and  becomes more prominant as x increases. From Fig. 4 and 6  it can be inferred that  metal insulator transition occurs when $\left<r_{A}\right>$ exceeds a critical value $\it{r_{MI}}\sim$ 1.314 at x = 0.2. The Mn-O bond lengths are almost equal
at $\left<r_{A}\right>$ $\gg$ 1.314 where metallicity is observed. This can be due to the removal of static coherent distortion of MnO$_{6}$ octahedra which causes charge localization in the metallic phase$^{\ref{rada}}$.\\

The direct exchange interaction between Mn ions is not possible in  manganites because of the intervening oxygen. According to Goodenough$^{\ref{go}}$, the nature of interaction depends on the extent of overlap between Mn d-orbitals and O p-orbitals which is very sensitive to Mn-Mn distance. In the Mn-O-Mn bonding patterns, if both Mn-O bonds are covalent the Mn-Mn distance will be shortest and interaction will be antiferromagnetic (AFM). If one Mn-O bond is covalent and the other ionic, the Mn-Mn separation is larger and results in ferromagnetic (FM) interaction. Magnetic properties of NdMnO$_{3}$ where in-plane FM and out-of-plane AFM interactions have been noticed$^{\ref{tok},\ref{go}}$ are consistent with this theoretical prediction. However, the predominant mechanism involved in doped NdMnO$_{3}$ is Double Exchange (DE) ferromagnetic interaction. A.C susceptibility data shown in Fig.5 clearly exhibits the presence of FM phase at low temperature in the range of x of the present investigation. In doped systems, Mn$^{4+}$-O-Mn$^{3+}$ bonding arrangement is degenerate with Mn$^{3+}$-O-Mn$^{4+}$. This implies charge hopping between Mn ions leading to metallicity which is observed in the present case at x $\gg$ 0.2(Fig.4). In DE, Mn-Mn distance is shorter than that of the other feromagnetic case (described before) and hopping integral between Mn d-orbitals is proportional to cos($\theta_{ij}$/2) where $\it{\theta_{ij}}$ is the angle between two Mn moments at i and j sites $^{\ref{ph}}$. Since, the direction of the Mn moment relative to the axes of the MnO$_{6}$ octahedron is determined in large part by crystal field, it is expected that $\it{\theta_{ij}}$ will depend on the Mn-O-Mn tilt angle$^{\ref{jf}}$. So, the hopping rate will depend on Mn-O-Mn bond angle $^{\ref{jf}, \ref{coey}, \ref{hy}}$ which in turn will depend on the magnitude of tilt between two neighbouring octahedra. The deviation of Mn-O-Mn bond angle from 180$^{0}$  is  maximum  in  NdMnO$_{3}$ and  decreases with increase of x (Fig.7).The Mn-Mn bond distances and corresponding magnetic interactions for various dopant concentration are shown in Table.6. Thus, it is evident that structural changes due to  doping plays an important role to control the physical properties of the system.  

\section*{Conclusions}
 The structure of Nd$_{1-x}$Pb$_{x}$MnO$_{3}$ single crystals are deterimined for x = 0.25 and 0.37. Structural changes  from orthorhombic $\rightarrow$ tetragonal $\rightarrow$ cubic are discernible with increasing dopant concentration (x). No evidence for twinning is observed in electron diffraction experiments. The MnO$_{6}$ octahedral distortion and inter octahedral tilt are  removed  progressively with higher doping.
 A correlation between the modulation in structure and physical  properties has been presented. It is observed that, insulator to metal as well as paramagnetic to ferromagnetic transition temperatures are  sensitive to divalent doping concentration.

\section*{Acknowledgements}  

This work is supported by CSIR  through an extramural research grant which is gratefully acknowledged. We thank the Department of Science and Technology, India for data collection on the CCD facility setup under the IRPHA-DST program.

\section*{References}

\begin{enumerate}

\item\label{gill} M. A. Gilleo, $\it{ Acta\ Cryst. }$, 1957, $\bf{10}$, 161.
\item\label{woo} W. C. Yi,  S. I. Kwun and  J. G. Yoon, $\it{J.\ Phys.\ Soc.\ Japan}$, 2000,
$\bf{69}$, 2706.
\item\label{poll} E. Pollert, S. Krupicka and E. Kuzmicova, $\it{J.\ Phys.\ Chem.\ Solids}$, 1982, $\bf{43}$, 1137. 
\item\label{pg} P. G. Radaelli, M. Marezio, H. Y. Hwang  and  S. W. Cheong, $\it{J.\ Solid\ State\ Chem.}$, 1996, $\bf{122}$, 444.
\item\label{dn} D. N. Argyriou, D. G. Hinks, J. F. Mitchell, C. D. Potter, A. J. Schultz, D. M Young, J. D. Jorgensen and  S. D. Bader, $\it{J.\ Solid\ State\ Chem.}$, 1996, $\bf{124}$, 381.
\item\label{jf} J. F. Mitchell, D. N. Argyriou, C. D. Potter, D. G. Hinks, J. D. Jorgensen and S. D. Bader, $\it{Phys.\ Rev.\ B}$, 1996, $\bf{54}$, 6172.
\item\label {jt} J. Topfer and  J. B. Goodenough, $\it{J. Solid\ State\ Chem.}$, 1997, $\bf{130}$, 117.
\item\label {Urushi}  A. Urushibara, Y. Moritomo, T. Arima, A. Asamitsu, G. Kido and Y. Tokura, $\it{Phys.\ Rev.\ B}$, 1995, $\bf{51}$, 14103.

\item\label{kaji} R. Kajimoto, H. Yoshizawa, H. Kawano, H. Kuwahara, Y. Tokura, K. Ohoyama and  M. Ohashi, $\it{Phys.\ Rev.\ B}$, 1999, $\bf{60}$, 9506.

\item\label{noel} N. W. Thomas, $\it{Acta\ Cryst.}$, 1998, $\bf{B54}$, 585.

\item\label{rao} C. N. R. Rao and  A. K. Raychaudhuri in \\ 
$\it{Colossal\ Magnetoresistance,\ Charge\ Ordering\ and\ Related\ Properties\ of\ Manganese\ Oxides}$, ed. C.N. R. Rao and  B. Raveau, World Scientific, Singapore, 1998, p.3.
\item\label{tok} J. F. Mitchell, D. N. Argyriou and J. D. Jorgensen in\\ $\it{Colossal\ Magnetoresistive\ Oxides}$, ed. Y. Tokura,  Gordon $\&$ Breach Science Publishers, New York, 2000, p. 189.

\item\label{jb} J. B. Torrance, P. Lacorre, A. I. Nazzal, E. J. Ansaldo and   Ch. Niedermayer,  $\it{Phys.\ Rev.\ B}$,  1992, $\bf 45$,  8209.
\item\label{good} J. B. Goodenough in $\it{Structure\ and\ Bonding,\  Vol.\ 98}$, ed. J.B. Goodenough, Springer-Verlag Berlin, Heidelberg, 2001, p.7.
\item\label{kana} J. Kanamori, $\it{J.\ Phys.\ Chem.\ Solids}$, 1959, $\bf{10}$, 87.
\item\label{coey} J. M. D. Coey, M. Viret and S. von Molnar, $\it{Adv.\ Phys.}$, 1999, 
$\bf{48}$, 167. 

\item\label{nil} N. Ghosh, S. Elizabeth, H.L. Bhat, G.N. Subanna and  M. Sahana, $\it{ J.\ Magn.\ Magn.\ Matter.}$, 2003, $\bf{256}$, 286.

\item\label{bruk} Bruker, SMART and SAINT, Bruker AXS Inc., 1998, Madison, Wisconsin, USA.
\item\label{shel} G. M. Sheldrick, SHELXL-97, 1997, University of Gottingen, Germany.
\item\label{faal} S. Faaland, K.D. Knudsen, M.-A. Einarsrud, L. Rormark, R. Hoier and T. Grande, $\it{J.\ Solid\ State\ Chem.}$, 1998, $\bf{140}$,320.
\item\label{hl} H. L. Ju, J. Gopalakrishnan, J. L. Peng, Qi Li, G. C. Xiong, T. Venkatesan  and  R. L. Green, $\it{Phys.\ Rev.\ B}$, 1995, $\bf{51}$, 6143.
\item\label{alek} K. S. Aleksandrov, $\it{Ferroelectrics}$, 1976, $\bf{14}$, 801.
\item\label{bar} H. Barnighausen., $\it{Acta\ Cryst.}$, 1975, $\bf{A31}$, S31.
\item\label{Mun} A. Munoz, J. A. Alonso, M. J. Martinez-Lope, J. L. Garcia Munoz  and  M. T. Fernandez-Diaz, $\it{J.\ Phys. : Condens.\ Matter}$, 2000, $\bf{12}$, 1361.
\item\label{hy} H. Y. Hwang, S. W.Cheong, P. G. Radaelli, M. Marezio and  B. Batlogg,
 $\it{Phys.\ Rev.\ Lett.}$, 1995, $\bf{75}$, 914.
\item\label{rada} P. G. Radaelli, M. Marezio, H. Y. Hwang, S. W. Cheong and  
B. Batlogg, $\it{Phys.\ Rev.\ B}$, 1996, $\bf{54}$, 8992.

\item\label{go} J. B. Goodenough, $\it{Phys.\ Rev.}$, 1955, $\bf{100}$, 564.
\item\label{ph} P. W. Anderson  and  H. Hesegawa, $\it{Phys.\ Rev.}$, 1955, $\bf{100}$, 675.


\end{enumerate}


\begin{table}
\caption{ Crystal data, Measurements and Refinement parameters for
 Nd$_{0.75}$Pb$_{0.25}$MnO$_{2.72}$ and Nd$_{0.57}$Pb$_{0.37}$MnO$_{3}$}
\begin{center}
\begin{tabular}{|c|}\hline
\bf{Crystal Data} \\ \hline
\end{tabular}
\end{center}

\begin{center}

\begin{tabular}{|c|c|c|}\hline
Empirical-Formula &  Nd$_{0.75}$Pb$_{0.25}$MnO$_{2.72}$ & Nd$_{0.57}$Pb$_{0.37}$MnO$_{3}$\\ \hline 
Crystal Symmetry   &  Tetragonal &  Cubic\\ \hline
Crystal habit, colour & Blocks, black & Blocks, Black\\ \hline
Crystal Size(mm) & 0.3 x 0.2 x 0.04 mm$^{3}$ & 0.18 x 0.057 x 0.2 mm$^{3}$
\\ \hline
Crystal System  & Tetragonal & Cubic\\ \hline
Space group  &  P4/mmm  & Pm$\overline{3}$m\\ \hline
Cell dimensions($\AA$) & a = b = 7.725(1), c = 3.884(1) & a = b = c = 7.737(2)\\ \hline
                                          
Volume($\AA^{3}$) &  234.19(7) & 463.22(2)\\ \hline
Formula Weight    & 263.09 & 697.8 \\ \hline
Density(calculated) & 7.457g/cm$^{3}$ & 6.73g/cm$^{3}$\\  \hline
Z & 4 & 8\\ \hline
F(000) & 457.8 &  817.7\\ \hline 
\end{tabular}
\end{center}
\end {table}

\begin{table}
\begin{center}
\begin{tabular}{|c|}\hline
\bf{Data Collection}\\ \hline
\end{tabular}
\end{center}
\begin{center}
\begin{tabular}{|c|c|c|}\hline
Equipment & Bruker APEX  & Bruker APEX \\
          & SMART CCD &  SMART CCD \\  \hline
$\lambda$(Mo K$\alpha$ & 0.7107 & 0.7107\\ (graphite &   & \\
monochromator))(\AA)&  &  \\  \hline

Scan mode & $\omega$ scan & $\omega$ scan \\ \hline
Temeprature(K) & 298 & 298 \\ \hline

$\theta$ range(deg) & 2.8-28 & 2.6-27.5\\ \hline

Recording  & -10$\leq h\leq$ 9, & -10$\leq h\leq$ 9, \\
reciprocal  &  -10$\leq k \leq$ 9 & -9$\leq k \leq$ 9 \\
space & -5$\leq l \leq$ 5 & -9$\leq l \leq$ 9 \\  \hline

Number of measured & 1336 & 3621\\
 reflections  &   & \\ \hline
Number of independent  & 202[R(int)= 0.0443] & 143[R(int)= 0.0489] \\ 
reflections &   &   \\ \hline
$\mu$(mm-1) &  39.485 & 33.096\\ \hline
\end{tabular}
\end {center}
\end {table}

\begin{table}
\begin{center}
\begin{tabular}{|c|}\hline
\bf{Refinement}\\ \hline
\end{tabular}
\end{center}

\begin{center}
\begin{tabular}{|c|c|c|}\hline
Number of refined parameters & 18 & 18\\  \hline
Refinement method  & Full matrix least squares & Full matrix least squares \\ \hline
R[I$\succ$ 4$\sigma$I]/R[all data] & 0.083/0.087 & 0.103/0.101\\ \hline
WR[I$\succ$ 4$\sigma$I]/R[all data] & 0.233/0.226 & 0.394/0.39\\ \hline
GoF  & 1.323 & 2.021\\  \hline
Max/min $\Delta$ $\rho$ \AA e$^{-3}$ & 3.987/-3.448 & 5.318/-8.343
\\  \hline
\end{tabular}
\end {center}
\end{table}

\begin{table}
\caption{Fractional atomic coordinates and anisotropic thermal parameters (U$_{ij}$) at 298K for Nd$_{0.75}$Pb$_{0.25}$MnO$_{2.72}$}
\begin{center}
\begin{tabular}{|c|c|c|c|c|c|c|}\hline
Atom & X & Y & Z & site & Occupancy & U$_{equvi}$\\ \hline
Nd1  & 0.5 & 0 & 0.5 & 2e & 1.0 & 0.0322(12)\\ \hline
Pb1   &  0.5 & -0.5  & 0.5 & 1d & 1.0 & 0.02444(10)\\ \hline
Nd2 & 0 & 0 & -0.5 & 1b & 1.0 & 0.0730(26)\\ \hline
Mn1  & 0.2504(4)& -0.2504(4)& 0 & 4j & 1.0 & 0.0261(16)\\ \hline
O1  & 0.2562(6) & -0.2562(6) & -0.5 &  4k & 0.96 & 0.0926(241)\\ \hline
O2 & 0 & -0.2508(7)& 0 & 4l & 1.0 & 0.1002(215)\\ \hline
O3 & 0.2655(9) & -0.5 & 0.0 & 4n & 0.80 & 0.1009(268)\\ \hline
\end{tabular}
\end{center}
\end {table}

\begin{table}
\caption{Fractional atomic coordinates and anisotropic thermal parameters (U$_{ij}$) at 298K
 Nd$_{0.57}$Pb$_{0.37}$MnO$_{3}$}
\begin{center}
\begin{tabular}{ |c|c|c|c|c|c|c| } \hline
Atom & X & Y & Z & site & Occupancy & U$_{equvi}$\\ \hline
Nd1  & 0 & 0 & 0 & 1a & 1.0 & 0.015(3)\\  \hline
Nd2  & 0 & 0.5  & 0.5 & 3c & 0.96 & 0.041(3)\\ \hline
Nd3  & 0.5 & 0.5 & 0.5 & 1b & 0.72 & 0.060(6)\\ \hline
Pb1  &  0 &  0.5 & 0 & 3d & 1.0 & 0.034(2)\\ \hline
Mn1  & 0.2499(4)& 0.2499(4) & 0.2499(4) & 8g & 1.0 & 0.030(7)\\ \hline
O1  & 0.2460(40) & 0.2460(40) & 0.5 & 8g  & 1.0 & 0.120(5)\\ \hline
O2 & 0.2520(30) & 0.2520(30) & 0 &  12i & 1.0 & 0.110(40)\\ \hline
\end{tabular}
\end{center}
\end{table} 

\begin{table}
\caption{ Selected bond lengths(\AA) and bond angles ($^{0}$) for Nd$_{0.75}$Pb$_{0.25}$MnO$_{2.72}$}
\begin{center}
 
\begin{tabular}{|c|}\hline
4x(Nd1-O1) = 2.74(3)\\ \hline
4x(Nd1-O2) = 2.749(1)\\ \hline
4x(Nd1-O3)= 2.832(1)\\ \hline
4x(Nd2-O1)= 2.81(3)\\ \hline
4x(Nd2-O2)=2.75(3)\\ \hline
4x(Pb1-O1) = 2.67(3)\\ \hline
4x(Pb1-O3) = 2.66(4)\\ \hline
\end{tabular}             
\end{center}
\end{table}   

\begin{table}
\begin{center}
\begin{tabular}{|c|c|}\hline

Mn1-O1 = 1.943(2) & O1-Mn1-O1 = 176(3) \\ \hline
Mn1-O1 = 1.943(2) & O3-Mn1-O2 = 176(3)\\ \hline
Mn1-O2 = 1.944(4) & $\cdots$ \\ \hline 
Mn1-O3 = 1.942(7) & Mn1-O1-Mn1 = 176(2) \\ \hline
Mn1-O3 = 1.942(7) & Mn1-O2-Mn1 = 179(3) \\ \hline
Mn1-O2 = 1.944(4) & Mn1-O3-Mn1 = 173(4) \\ \hline
\end{tabular}
\end{center}
\end{table}

\begin{table}
\caption{ Selected bond lengths(\AA) and bond angles($^{0}$) for Nd$_{0.57}$Pb$_{0.37}$MnO$_{3}$}
\begin{center}
\begin{tabular}{|c|}\hline

12x(Nd1-O2) = 2.76(5)\\ \hline
8x(Nd2-O1) = 2.736(1)\\ \hline
4x(Nd2-O2) = 2.71(5) \\ \hline
4x(Pb1-O1) = 2.69(4) \\ \hline
8x(Pb1-O2) = 2.736(1) \\ \hline
3x(Mn1-O1) =  1.933(3)\\ \hline
1x(Mn1-O2) = 1.933(3) \\ \hline
2x(Mn1-O2) = 1.934(3) \\ \hline
\end{tabular}
\end{center}
\end{table}

\begin{table}
\begin{center}
\begin{tabular}{|c|}\hline
O1-Mn1-O1 = 91.7(1)\\ \hline 
O1-Mn1-O2 = 179.6(1)\\ \hline
O1-Mn1-O2 = 88.6(1)\\ \hline
Mn1-O1-Mn1 = 178(3)\\ \hline
Mn1-O2-Mn1 = 178(3)\\ \hline
\end{tabular}             
\end{center}
\end{table}

\begin{table}
\caption{Mn-Mn distances and corresponding interactions in pure and doped NdMnO$_{3}$(x = 0)$^{\ref{Mun}}$ structures.}
\vspace{1cm}
\begin{center}
\begin{tabular}{|c|c|c|}\hline
Formula  &  Mn-Mn distance($\AA$) &  Interaction \\ \hline
NdMnO$_{3}$  &  3.7945(6)  &  AFM(out of plane) \\ 
             &   3.9342(4) &  F\ M\ (\ in\ plane\ )\\ \hline
Nd$_{0.75}$Pb$_{0.25}$MnO$_{2.72}$ &  3.8836, 3.8878, 3.876  & F M (DE)\\
             &           &      \\ \hline
Nd$_{0.57}$Pb$_{0.37}$MnO$_{3}$ & 3.8654, 3.8674  & F M (DE) \\ 
              &           &     \\ \hline

\end{tabular}
\end{center}
\end{table}

\section*{Figure Captions}

\noindent {Fig.1}\
  Packing diagram of (a) Nd$_{0.75}$Pb$_{0.25}$MnO$_{2.72}$
and   (b) Nd$_{0.57}$Pb$_{0.37}$MnO$_{3}$  along the b-direction.\

\noindent {Fig.2(a)}\ 
The simulated and experimental powder x-ray diffraction patterns for x = 0.25.\

\noindent {Fig.2(b)}\ 
The simulated and experimental powder x-ray diffraction patterns for x = 0.37.\

\noindent {Fig.3}\
Electron diffraction pattern of Nd$_{1-x}$Pb$_{x}$MnO$_{3}$ for (a) x = 0.25 and (b) x = 0.37 respectively.\

\noindent {Fig.4}\
  Resistivity vs temperature plots  of  Nd$_{1-x}$Pb$_{x}$MnO$_{3}$ for x = 0.2, 0.3 and 0.5 in the range of 70-300 K.  \

\noindent {Fig.5}\
The A.C susceptibility plots for  x = 0.2, 0.3 and 0.5 at  100Hz and magnetic field $\sim$ 5 Oe. \

\noindent {Fig.6}\
  Mn-O bond lengths  as a function of lead concentration. Inset:- Mn-O bond lengths as a function of $\left<r_{A}\right>$. The  points indicate experimental data and the  lines are guide to the eye.  The data for x = 0 has been taken from Ref \ref{Mun}. \

\noindent {Fig.7}\
Tolerance factor (t$_{G}$) and  $\delta\theta$  plotted as a function of Pb concentrations(x), where $\delta\theta$ = 180 - $\theta_{ij}$. The  points indicate experimental  values and  line guide  to the eye. The data for x = 0 has been taken from Ref \ref{Mun}. \

\end{document}